# Generation of pseudo-high-order group velocity locked vector solitons in fiber lasers

XINXIN JIN,[1] ZHICHAO WU,[2] LEI LI,[1] YANQI GE,[1] JIAOLIN LUO,[1] QIAN ZHANG,[1] DINGYUAN TANG,[1] DEYUAN SHEN,[1] SONGNIAN FU,[2] DEMING LIU,[2] AND LUMING ZHAO[1],*

[1]Jiangsu Key Laboratory of Advanced Laser Materials and Devices, School of Physics and Electronic Engineering, Jiangsu Normal University, Xuzhou 221116, China
[2]Next Generation Internet Access National Engineering Lab (NGIA), School of optical and electronic information, Huazhong University of Sci&Tech (HUST), 1037 Luoyu Road, Wuhan 430074, China
*Corresponding author: lmzhao@ieee.org

**We propose and experimentally demonstrate the generation of pseudo-high-order group velocity locked vector solitons (GVLVS) in a fiber laser using a SESAM as the mode locker. With the help of an external all-fiber polarization resolved system, a GVLVS with a two-humped pulse along one polarization while a single-humped pulse along the orthogonal polarization could be obtained. The phase difference between the two humps is 180°.**

**Keywords:** Lasers, fiber;  Mode-locked lasers;  Pulse propagation and temporal solitons.

Solitons, as stable localized nonlinear waves, have been extensively studied. It is known that the dynamics of generated soliton pulse in single mode fiber (SMF) is determined by the nonlinear Schrödinger equation (NLSE). Generally, a SMF always supports two orthogonal polarization modes, which provides the possibility for the generation of vector solitons [1, 2]. According to the strength of fiber birefringence, various types of vector solitons, such as the phase-locked vector solitons (PLVSs) [3], the polarization rotating vector solitons [4], and the group velocity locked vector solitons (GVLVSs) [1], can be generated in a SMF. In terms of GVLVS, the central wavelength of orthogonal polarizations could have a shift, which can compensate the group velocity difference induced by the fiber birefringence. Thus the two pulses along orthogonal polarizations can be trapped together and co-propagate as a non-dispersive unit [5]. Apart from optical fibers, vector solitons have also been observed in mode-locked fiber lasers [3, 5-7]. Pulse formation mechanism in a fiber laser cavity is quite different from that in a SMF. Apart from the mutual interaction between group velocity dispersion (GVD) and nonlinear Kerr effect, soliton generation in fiber lasers is also determined by the cavity gain and loss, as well as the cavity boundary condition. When the cavity is operated in the anomalous dispersion regime, soliton sidebands are generated on the optical spectrum due to the periodic discrete perturbations of the cavity components. Previous investigation on the traditional soliton sidebands prove that their occurrence is pair-wise symmetric to the central wavelength, while their position at the spectrum is governed by the soliton parameters and cavity dispersion. As a result, the sidebands can be utilized as an indicator of wavelength shift between two soliton pulses at orthogonal polarizations. In 1988, Christodoulides and Joseph first theoretically predicted a high-order PLVS [8]. Then, stable high-order PLVS has been experimentally observed at the output of a mode-locked fiber laser [9], where one of the vector soliton components has a double-humped structure with 180° phase difference between two humps, while the orthogonal component has single-hump. To the best of our knowledge, no high-order GVLVS has been demonstrated.

In this paper, we propose an approach to generate a stable pseudo-high-order GVLVS based on the fundamental GVLVS which could be directly obtained in a passively mode-locked fiber laser. By passing the fundamental GVLVS through a polarization controller with appropriate phase retard, it is feasible to obtain a high-order GVLVS which has a polarization component consisting of two humps with 180° phase difference. As the high-order GVLVS is not directly generated from the fiber laser, we call it "pseudo-high-order" GVLVS. Experimental observations well demonstrate our proposal.

As shown in Fig. 1, a general GVLVS has two wavelength-shifted, single-humped pulses along different polarizations. By rotating the PC, we can simultaneously change the orientation of the GVLVS and the phase difference between the orthogonal components. After traversing the PC, The two polarizations of the GVLVS can be characterized by

$$F_1 = A_1 \operatorname{sech}(1.763(t - \Delta T/2)/T_0) \exp[i(2\pi c t/\lambda_1)]. \quad (1)$$

$$F_2 = A_2 \operatorname{sech}(1.763(t + \Delta T/2)/T_0) \exp[i(2\pi c t/\lambda_2 + \varphi_t)]. \quad (2)$$

where we assume the polarization components of the GVLVS both have Sech² profile. $A_1$ and $A_2$ are the amplitudes of the optical pulses along the two orthogonally polarizations. $\Delta T$ is the time separation between the two components. $\lambda_1$ and $\lambda_2$ are the central wavelength. $\varphi_t = \varphi_i + \Delta\varphi$, in which $\varphi_i$ represents the initial phase difference and $\Delta\varphi$ is the phase difference caused by the PC. An inline polarization beam splitter (PBS) is connected to the PC for seperating the orthogonal polarizations. The angle between $F_1$ and the horizontal axis is $\theta$. The spectral intensity and pulse intensity of the horizontal and vertical axis after PBS are calculated. Typical numerical results are shown in Fig. 2. We used the following parameters: $A_1 = 25, A_2 = 10$,

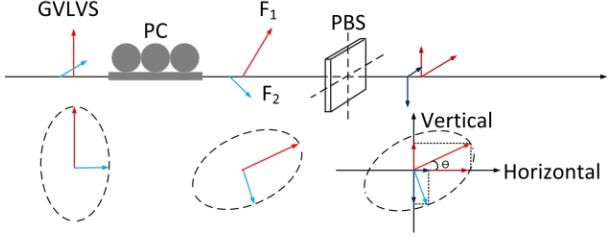

Fig. 1. Schematic of polarization evolution of a GVLVS when the input GVLVS goes through a PC and an inline PBS. PC, polarization controller; PBS, polarization beam splitter.

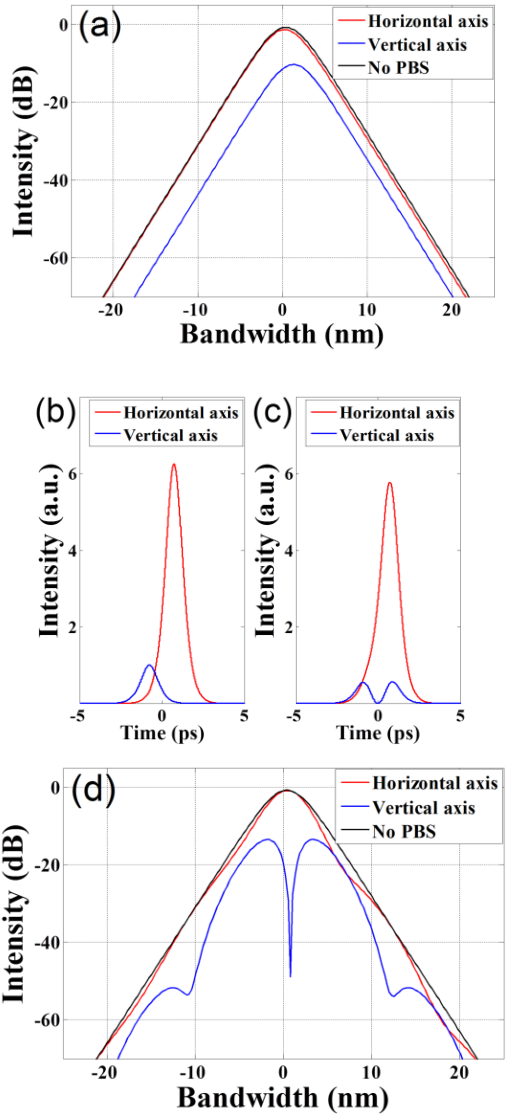

Fig. 2. (a) Optical spectrum and (b) temporal profile of the numerically obtained wavelength shift operation ($\theta = 0°$). (c) temporal profile and (d) optical spectrum of the numerically obtained pseudo-high-order operation ($\theta = 22°$, $\varphi_t = 0°$).

$\Delta T = 1.5$ ps, $T_0 = 1.2$ ps, $\lambda_1 = 1574$ nm, $\lambda_2 = 1575$ nm. Numerically we found that when $\theta = 0°$, the components along the horizontal and vertical axis always present obvious wavelength shift regardless of the value of $\varphi_t$ [Fig. 2 (a)]. The calculated pulse intensity profiles along the horizontal and vertical axis are both single-humped [Fig. 2 (b)]. In this case, the two orthogonal components of the GVLVS are completely resolved to horizontal and vertical direction after the PBS. Numerically we found that, when $\varphi_t = 0$, $\theta=22°$, a GVLVS with a two-humped pulse along one polarization while a single-humped pulse along the orthogonal polarization could be obtained [Fig. 2 (c)]. We note that the single-humped pulse is actually superposed by two pulses with same phase. The optical spectra of the components are shown in Fig. 2(d). There is a strong spectral dip at the center of the spectrum of the weaker component. Based on the pulse intensity profiles numerically calculated, obviously the spectral dip is formed due to the spectral interference between the two humps, and the strong dip at the center of the spectrum indicates that the phase difference between the two humps is 180°. The state seems to be a high order GVLVS. However, it is obtained by tuning the external PC before the external PBS. Therefore we call it a pseudo-high-order GVLVS. We note that high order solitons along both two polarizations could be obtained if the time separation between the two orthogonal components of the fundamental GVLVS is large enough.

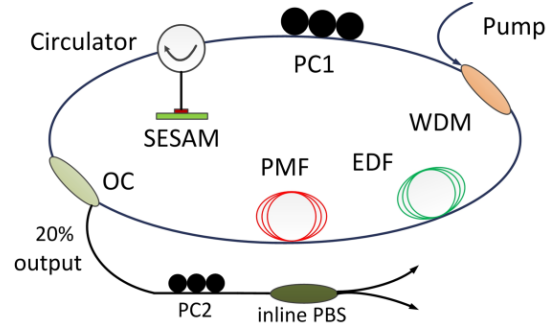

Fig. 3. Experimental setup of proposed fiber laser: WDM, wavelength-division multiplexer; EDF, Er-doped fiber; PMF, polarization-maintaining fiber; OC, optical coupler; PC, polarization controller; PBS, polarization beam splitter.

Encouraged by the numerical simulations, we then further searched for the pseudo-high-order GVLVS experimentally. The proposed fiber laser is schematically showed in Fig. 3. The self-starting mode-locking is achieved by a commercial SESAM, which is butt-coupled to the end surface of a standard FC/PC fiber connector. The SESAM possesses low-intensity absorption of 8% at the wavelength of 1550 nm, modulation depth of 5.5%, and relaxation time of 2 ps. The circulator is used to incorporate the SESAM into the cavity, as well to guarantee unidirectional propagation and suppress detrimental reflections. A 2 m EDF with a group velocity dispersion (GVD) parameter of ~ -48 ps/nm/km acts as the gain medium, which is pumped by a 1480 nm laser diode (LD) with maximum output power of 1 W. To avoid residual pump power damaging the SESAM, the pump beam is coupled into the cavity with the back-pumping scheme. A 0.25 m polarization-maintaining fiber (PMF) is used to enhance the cavity birefringence. The fiber pigtails of optical components are 11.75 m standard SMF with GVD parameter of 17 ps/nm/km. Thus, the ring cavity is dispersion-managed and operated in the anomalous dispersion region. The total optical cavity length is estimated to be around 14.4 m. The PC is used to finely tune the net cavity birefringence. A 20:80 fiber optical coupler (OC) is utilized as the output port of the fiber laser. Along the output port, another PC together with an inline PBS can seperate the orthogonal polarizations.

With appropriate setting of PC1, self-started mode-locking can be obtained easily by increasing the pump power above the mode-locking threshold of 140 mW. In contrast to the nonlinear polarization rotation (NPR) method, SESAM based mode-locking is polarization-insensitive, therefore under suitable condition of the cavity birefringence, vector solitons can always be observed. Figure 4(a) shows the optical spectra of the generated GVLVSs with two sets of soliton sidebands. To instigate the characteristics of extra set of spectral sidebands, we further monitor the optical spectrum and the pulse trains of the GVLVS along different orthogonal polarizations. Through rotating PC2 outside the cavity, we can simultaneously change the directions of two orthogonally polarized components. With the help of a PBS, two orthogonal polarizations are able to project to horizontal and vertical direction, respectively. After the polarization resolved measurement of the GVLVS, the components along each of two orthogonal directions present obvious wavelength shift. Due to the PBS limited polarization extinction ratio, there exists little residual light from the orthogonal axis, so that we can observe weak intensity at the position of sidebands belong to the other axis. Figure 4(b) shows the polarization resolved pulse trains trapping each other and co-propagating as a non-dispersive unit. Figure 4(c) shows the autocorrelation trace of the vector soliton. The pulse-widths of the two components are both ~1.16 ps if a sech$^2$ profile is assumed. Further rotating PC2, we search for the maximum pulse intensity at one of PBS output ports and record the soliton spectra of two orthogonal components, as shown in Fig. 5(a). The spectra have the same central wavelength and about 12 dB peak intensity difference. Both spectra possess two sets of soliton sidebands. Comparing to the fundamental vector solitons, there is an evident spectral dip at the central wavelength of the vertical component. No such dip appears in the spectrum of horizontal component. Figure 5(b) shows the autocorrelation traces. The vertical component has a double-humped intensity profile. The pulse-width of the humps is about 932 fs, and the separation between the humps is about 1.76 ps. The horizontal component has a pulse-width of 1.20 ps. The state we obtained is exactly a high-order GVLVS, where two humps along one polarization have out-of-phase relationship while there is only one hump along the orthogonal polarization.

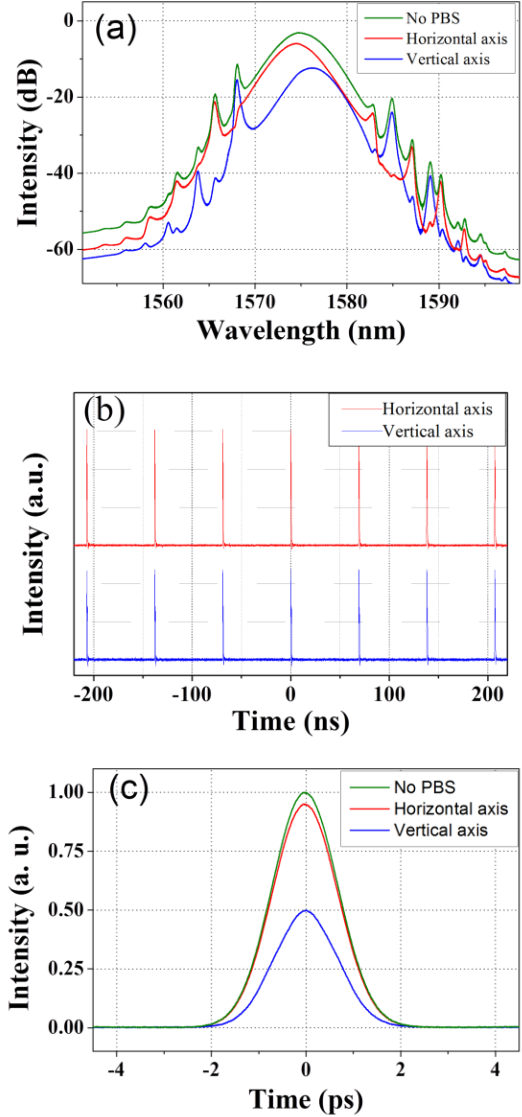

Fig.4. Operation of wavelength shift. (a) optical spectra. (b) oscilloscope trace. (c) autocorrelation trace.

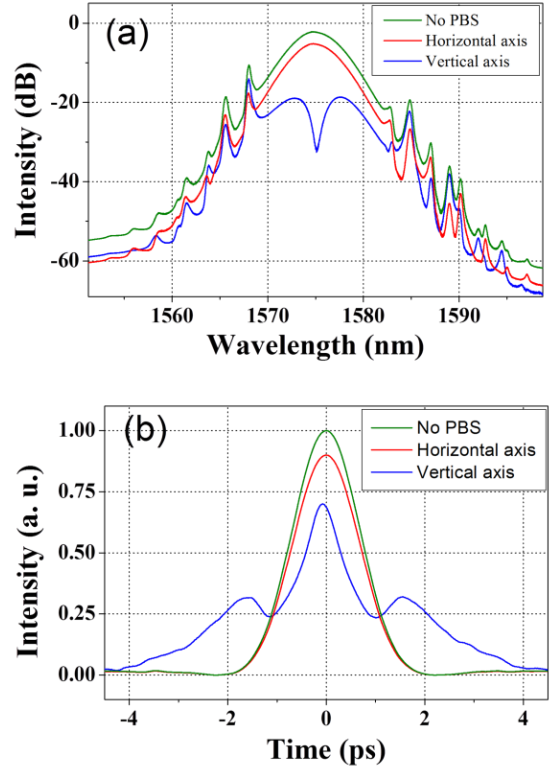

Fig. 5. Operation of pseudo-high-order vector soliton. (a) peak-dip optical spectrum. (b) double-humped autocorrelation trace.

Experimentally the position of the spectral dip of the vertical component could be shifted by rotating the PC2, which agrees well with the numerical simulations. The spectral peak intensity difference of ~12 dB matches with the numerical simulation. We note that to monitor the autocorrelation trace of the vertical component which is much weaker, the polarization-resolved pulse was amplified before being directed into the autocorrelator. Therefore, the pulse intensity difference from Fig. 5(b) did not show the real pulse intensity difference.

In conclusion we propose a method to generate high-order GVLVS based on the fundamental GVLVS and experimentally demonstrate it in a fiber laser. By transmitting the fundamental GVLVS directly from the fiber laser through an external PC and with appropriate phase retard, it is feasible to transform the fundamental GVLVS into a high-order one. The high-order GVLVS is characterized by a spectral dip along one polarization, due to the interference between the two humps along one polarization. As it is not directly generated from the fiber laser, we consider it as the so-called "pseudo-high-order" GVLVS.

This work was supported in part by the National Natural Science Foundation of China (61275109, 61275069, 61331010), in part by the Priority Academic Program Development of Jiangsu higher education institutions (PAPD), in part by the Jiangsu Province Science Foundation (BK20140231), in part by the National Key Scientific Instrument and Equipment Development Project (No. 2013YQ16048702) and in part by a project funded by the Jiangsu Normal University for graduate students in research and innovation (KYLX_1427).